\documentclass[a4paper]{jpconf}
\usepackage{graphicx}
\usepackage{grffile}
\usepackage{epstopdf}
\usepackage{amsmath, amsthm}
\usepackage{float}    
\usepackage{enumerate}

\begin{document}
\title{A dynamical evolution of GRB-Afterglows in a new generic model}

\author{E Zouaoui and  N Mebarki}

\address{Laboratoire de Physique Math\'{e}matique et Subatomique,\\
 Fr\`{e}res Mentouri University, Constantine 1, Algeria}

\ead{esma.zouaoui213@gmail.com}

\begin{abstract}
A new model using a general expression of the radiation
energy and explaining  the dynamics of the afterglows is proposed.
It is shown that this model is suitable for the ultra-relativistic and non-relativistic phases as well as the study
of radiative and adiabatic fireballs.

\end{abstract}

\section{Introduction}

Untill now, the mystery of the gamma ray bursts (GRBs) which are the brightest and violent events in the universe after the Bing Bang is still a big puzzle for astrophysicists and observers to be understood \cite{1}. In fact, This phenomenon starts by a prompt emission of the GRBs where the energy $E\sim 10^ {51}- 10^{54} ergs$, than end up with remnant emission called afterglows. At this stage scientists were able to measure the redshift \cite{2} \cite{3} from the obtained spectral lines, showing their cosmological origin. So far, many works were done in the literature trying to simulate this phenomenon and understand it  \cite{4} \cite{5} \cite{6} \cite{7} \cite{8} \cite{9} \cite{10} \cite{11}. To be more specific the duality of fireball blast-wave model was successful where its predictions converge with data. In this model two jets explode from the core in two opposite directions, as a result they shock the surrounding medium and emit radiations where the wave lengths run from the gamma to the radio rays spectrum. In what follows we propose a GRB-afterglows hydrodynamical model taking  into account the synchrotron emission as a major radiation mechanism and neglecting the absorption (such as  synchrotron self-absorption) and diffusion (such as inverse Compton scattering) effects. It is very important to mention that the limitation of the earlier models \cite{25} is related to the compatibility with the Sedov solution starting from a non-relativistic phase where we have an adiabatic processes and surrounding interstellar medium (ISM). In the present paper, as it will be shown in the next sections, the Sedov solution was satisfied achieving the energy conservation in the adiabatic regime. In section 2, we present the conventional models \cite{12} \cite{13} \cite{15} then we introduce our generic new model. In section 3, we discuss our numerical results and finally in section 4, we draw our conclusions.

\section{Dynamics and radiation}

\subsection{Conventional dynamical models}
So far, in literature many models have been proposed to describe the expansion of the GRB remnants such as the one of Chiang et al of ref.\cite{12} where:
\begin{equation} \label{eq:1}
\frac{d\Gamma}{dm}=-\frac{\Gamma^{2}-1}{M}
\end{equation}
and
\begin{equation}
M=M_{0}+m+(1-\varepsilon)\frac{U}{c^{2}}
\end{equation}
here $\Gamma$, M, m, $\varepsilon$ and $dU=(\Gamma-1)dmc^{2}$ are the Lorentz  factor, the total mass of the fireball that includes the initial mass $M_{0}$ of the jet (ejected by the progenitor of the GRB), the swept mass from the external environment, the efficiency of radiation and the fraction of the internal non radiative energy, respectively. We remind that this first proposed model was written just for the relativistic phase expansion, because the solution of eq.\eqref{eq:1} is not consistent with Sedov results for an adiabatic expansion of non relativistic phase. On the other hand, Huang et al of ref.\cite{13} have proposed new generalized model which is more or less fairly good for ultra relativistic and non relativistic phases. In this ref.\cite{13}, the authors have used an infinitesimal difference equation for the internal non radiative energy such as $dU=d[(\Gamma-1)mc^{2}]$ which was assumed and proposed by Panaitescu et al \cite{14}. In this model, eq.\eqref{eq:1} is replaced by the following equation:
\begin{equation}\label{eq:2}
\frac{d\Gamma}{dm}=-\frac{\Gamma^{2}-1}{M_{0}+\varepsilon m+2(1-\varepsilon)\Gamma m}
\end{equation}
where the radiation efficiency $\varepsilon$ was assumed to be constant during the deceleration.

In 2002 Feng et al \cite{15} \cite{16} suggested that $\varepsilon$ is a variable belonging to the interval [0,1] and proposed an infinitesimal difference equation for U of the form:
\begin{equation}\label{eq:2343}
dU=d[(1-\varepsilon)U_{ex}]
\end{equation}
where $U_{ex}=(\Gamma-1)mc^{2}$ is the internal energy produced in the expansion. In this case one has \cite{17}:

\begin{equation}\label{eq:3}
\frac{d\Gamma}{dm}=-\frac{\Gamma^{2}-1}{M_{0}+m+U/c^{2}+(1-\varepsilon)\Gamma m}
\end{equation}
 with:
\begin{equation}\label{eq:33}
\varepsilon=\varepsilon_{e}\frac{{t^{^{\prime }}}^{-1}_{syn}}{{t^{^{\prime }}}^{-1}_{syn} + {t^{^{\prime }}}^{-1}_{ex}}
\end{equation}
Here
\begin{equation}\label{eq:31}
    t^{^{\prime }}_{syn}=6\pi m_{e}c / (\sigma_{T} B^{^{\prime }2}\gamma_{min})
\end{equation}
and
\begin{equation}\label{eq:32}
    t^{^{\prime }}_{ex}=R/(\Gamma c)
\end{equation}
are the synchrotron cooling and expansion times in the co-moving frame
respectively. The parameters $\varepsilon_{e}$  is a fraction of the internal
energy carried by the accelerated electrons in the jet, $m_{e}$ the
rest electron mass, $\sigma_{T}$  is the Thomson cross section and
$B^{^{\prime }}=(8\pi\varepsilon_{B}e{^{\prime }}){^{1/2 }}$  the
magnetic energy density where $\varepsilon_{B}$ is the fraction of
the kinetic energy of the shock converted into a magnetic energy and
$e{^{\prime }}$ an energy density defined in ref.\cite{18}. The
radius R of the blast-wave is determined by the relation \cite{12}:
\begin{equation}\label{eq:192}
\frac{dR}{dt}= \beta c \Gamma (\Gamma+\sqrt{\Gamma^{2}-1})
\end{equation}
where $\beta=v/c$ is the jet velocity and $\gamma_{min}$ the minimum Lorentz factor depending on the index p ($2<p<3$) and is given by \cite{18}:
\begin{equation}
\gamma_{min}=\varepsilon_{e}\bigg(\frac{p-2}{p-1}\bigg)\bigg(\frac{m_{p}}{m_{e}}\bigg)(\Gamma-1)+1
\end{equation}
$m_{p}$ being the rest proton mass. 

In the highly radiative case where $(\varepsilon\simeq1, U=0 )$,
eqs.\eqref{eq:2} and \eqref{eq:3}, together with eq.\eqref{eq:1}
reduce to the following differential equation \cite{13}:
\begin{equation}
\frac{d\Gamma}{dm}=-\frac{\Gamma^{2}-1}{M_{0}+m}
\end{equation}
where the analytic solution is \cite{13}:
\begin{equation}\label{eq:191}
\frac{(\Gamma-1)(\Gamma_{0}+1)}{(\Gamma+1)(\Gamma_{0}-1)}=\frac{(m_{0}+M_{0})}{(m+M_{0})}
\end{equation}

with $\Gamma_{0}$ and $m_{0}$ are the initial values of $\Gamma$ and $m$ respectively.

In the fully adiabatic case where  $(\varepsilon\simeq0, U=U_{ex})$, eqs.\eqref{eq:2} and \eqref{eq:3} become \cite{13}:
\begin{equation}\label{eq:18}
\frac{d\Gamma}{dm}=-\frac{\Gamma^{2}-1}{M_{0}+2\Gamma m}
\end{equation}
leading to the analytic solution \cite{13}:
\begin{equation}\label{eq:188}
(\Gamma-1)M_{0}c^{2}+(\Gamma^{2}-1)mc^{2}=E_{k0}
\end{equation}
where$E_{k0}$ is the initial value of the kinetic energy $E_{k}$.

\subsection{Our generic new model}

In the Feng et al model \cite{15}, and because of the behavior of the radiation efficiency $\varepsilon$ evolving with time, a differential equation for the internal residual energy  in the fireball for the dynamical evolution of the GRB-aftergolws was proposed. However the change in the radiative part was ignored.

In fact the total kinetic energy $E_{k}$ of the fireball is given by \cite{14}:
\begin{equation}\label{eq:5}
E_{k}=(\Gamma-1)(M_{0}+m)+\Gamma U
\end{equation}

The global conservation of energy implies that:
\begin{equation}\label{eq:6}
dE_{k}=-dE_{rad}
\end{equation}
where $dE_{rad}$ is the radiative part from the internal energy of the fireball. The definition given by Blandford et al is $ dE_{rad}=\varepsilon\Gamma(\Gamma-1)dmc^{2} $ \cite{19}. The internal energy created within the fireball is $U_{ex}$, such that:
\begin{equation} \label{eq:4}
dE_{rad}= \varepsilon \Gamma dU_{ex}
\end{equation}
This due to the fact that the internal energy can be radiative or remains in the fireball and has the same form up to a factor $ \varepsilon$ (resp. $ 1-\varepsilon$) for radiative (resp. residual) internal energy. Therefore, eq.\eqref{eq:4} can be rewritten as:
\begin{equation} \label{eq:20}
dE_{rad}= \varepsilon \Gamma [(\Gamma-1)dm+md\Gamma]c^{2}
\end{equation}

Using eqs. [\eqref{eq:5}- \eqref{eq:20}], one obtains the following differential equation describing the evolution of the fireball and consistent with the Sedov solution.
\begin{equation} \label{eq:7}
\frac{d\Gamma}{dm}=-\frac{\Gamma^{2}-1}{M_{0}+m(\Gamma+1)+U/c^{2}}
\end{equation}

It is important to notice that in the adiabatic case when $\varepsilon\simeq0$, eq.\eqref{eq:7} reduces to eq.\eqref{eq:18} while in the highly radiative case where $(\varepsilon\simeq1 )$ eq. \eqref{eq:7} gets the new form:
\begin{equation}\label{eq:190}
\frac{d\Gamma}{dm}=-\frac{\Gamma^{2}-1}{M_{0}+m(1+\Gamma)}
\end{equation}
leading to the following analytic solution:

\begin{equation}\label{eq:189}
\Gamma=-1+\frac{M_{0}}{m}productLog\bigg[\frac{m}{M_{0}}e^{\frac{2m-C}{M_{0}}}\bigg]
\end{equation}

(C is an integration constant)

Moreover, it was show that the total luminosity of the fireball has the following form:
\begin{equation}
L= L_{B}+\Gamma \varepsilon mc^{2}\frac{d\Gamma}{dt}
\end{equation}
where $L_{B}$ is the total luminosity used in the models of refs.\cite{15} \cite{13} and has as expression $L_{B}=\Gamma\varepsilon[(\Gamma-1)c^{2} dm/dt]$ and $\Gamma$ is giving by eq. \eqref{eq:189}.

\subsection{Synchrotron radiation}
The distribution of the accelerated electrons by the external shock behind the blast wave in the absence of the radiation loss is generally assumed to be a power-law function of the electron energy \cite{17}:
\begin{equation}
N^{\prime }_{e} (\gamma_{e})=\frac{dN^{\prime }_{e}}{d\gamma_{e}}=C^{te} \gamma^{-P}_{e},        \gamma_{min} \leq \gamma_{e} \leq \gamma_{max}
\end{equation}
where $\gamma_{max}=a 10^{7}(B^{^{\prime }}/1G)^{-1/2}$
 ref.\cite{19} is the maximum Lorentz factor and $"a"$ is a factor
taking its values between 1 and 10. The power for synchrotron
radiation $P_{\nu^{\prime }}$ is defined as \cite{18}:
\begin{equation}
P_{\nu^{\prime }}= \frac{2  \sqrt{3} e^{2}\nu_{L}}{c} \int^{\gamma_{max}}_{\gamma_{min}} N^{\prime }_{e} (\gamma_{e}) F\bigg(\frac{\nu^{\prime }}{\nu^{\prime }_{c}}\bigg) d\gamma_{e}
\end{equation}
where F(x) is the synchrotron function such that \cite{20}:
\begin{equation}
F(x)=x\int^{\infty}_{x} K_{5/3}(x^{^{\prime }}) dx^{^{\prime }}
\end{equation}
Here $K_{5/3}$ is the second kind modified Bessel function. The synchrotron frequency given $\nu^{\prime }_{c}$ by \cite{22}:
\begin{equation}
\nu^{\prime }_{c}=\frac{2}{3} \nu_{L}  \gamma^{2}_{e}
\end{equation}
where $\nu_{L}$ is the Lamor frequency:
\begin{equation}
\nu_{L}=\frac{1}{2 \pi}  \frac{e B^{\prime }}{2m_{e}c}
\end{equation}
Notice that the radiation in the lab frame can be calculated using the  relativistic transformations \cite{21} \cite{22}:
\begin{equation}
\nu=\frac{(1+\beta)\Gamma}{1+z} \nu^{^{\prime }}
\end{equation}
where z is the redshift.

The instantenous intensity $F_{\nu}$ giving the light curves at a frequency $\nu$ in Jansky ($erg.s^{-1}.cm^{-3}.Hz^{-1}$) is:
\begin{equation}\label{eq:8}
F_{\nu}=\frac{1}{4\pi D_{L}(z)^{2}} 4 \pi \frac{dP_{\nu}}{d\Gamma}
\end{equation}
where $D_{L}(z)$ is the luminosity distance in the $\Lambda CDM$ model with $\Omega_{M}=0.3$,$\Omega_{\Lambda}=0.7$ and $H_{0}=71 km.s^{-1}Mpc^{-1}$. Note that for observations by satellites or terrestrial telescopes one has to use $S_{B}$ by integrating $F_{\nu}$ in a giving interval $[\nu_{1},\nu_{2}]$:
\begin{equation}\label{eq:9}
S_{B}= \int^{\nu_{2}}_{\nu_{2}} F_{\nu}d\nu
\end{equation}

\section{Numerical results and discussion}

The variation of the mass swepted by the fireball during the deceleration is given by:
\begin{equation}
dm=4 \pi R^{2}nm_{p}dR
\end{equation}
where
\begin{equation}
dR= \beta c \Gamma(\Gamma+ \sqrt{\Gamma^{2}-1})dt
\end{equation}
here t is the observed time and measured by telescopes and satellite. In solving eqs. \eqref{eq:1}, \eqref{eq:2}, \eqref{eq:3} and \eqref{eq:7} numerically, the initial parameters are taken as: $\Gamma_{0}=250, M_{0}=2.10^{-6}M_{\odot}, p=2.1, \theta=10^{\circ}$ (solid angle parameters), $a =4.0$ and $g=0$ with $k=0, n=1cm^{-1}$ corresponding to the interstellar medium (ISM) ref.\cite{25}.

\begin{figure}[H]
	\begin{center}
		\includegraphics[width=30pc]{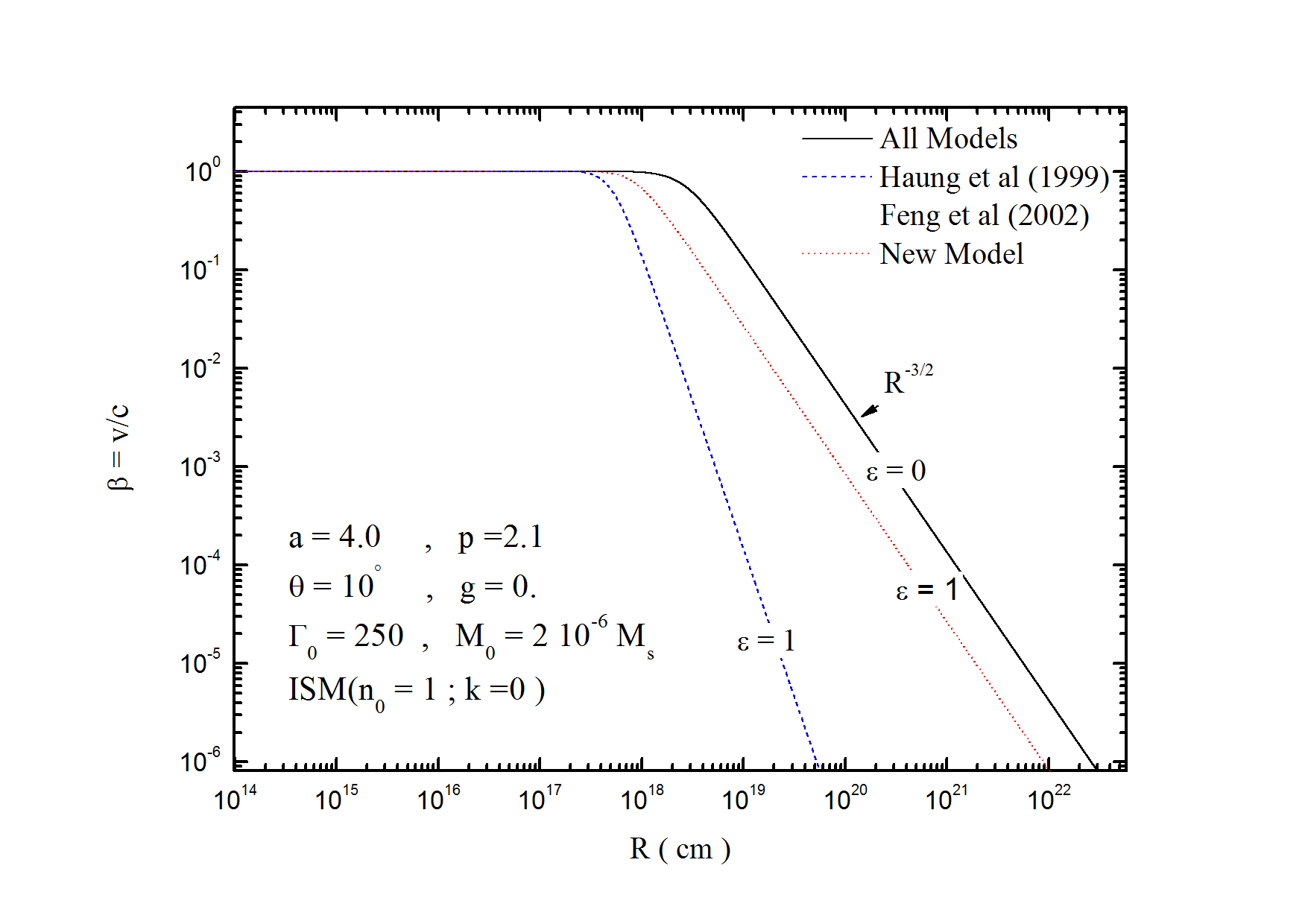}
	\end{center}
	\caption{\label{label}Evolution of the jet velocity $\beta=v/c$ as a function of the distance R (in logarithmic scale). For  radiative ($\varepsilon=1$) and adiabatic ( $\varepsilon=0$) cases.}
	\label{fig:fa1}
\end{figure}

Fig.\ref{fig:fa1} displays the evolution of the jet velocity calculated according to eqs. \eqref{eq:2}, \eqref{eq:3} and \eqref{eq:7} \cite{23}. Notice that the three models are compatible with the Sedov solution, in the non relativistic (NR) limit for the adiabatic case \cite{24}, where the velocity is proportional to $ R^{-3/2}$. However in the other regions the solution of eq.\eqref{eq:7} is  different from those of eqs. \eqref{eq:2}, \eqref{eq:3}. In fact for the ultra relativistic (UR) case where $ x=m/M_{0} << 1 $, the principal branch of Lambert productlog(z) function denoted by $W_{0}(z)$ with $z=xe^{2x-C}$ (C is integration constant) has the following series expansion near $\frac{1}{e}$ (e: exponential) as $W_{0}(x)\approx -1+\sqrt{2(e x+1)}+o(\sqrt{x+\frac{1}{e}})$ which leads to $\Gamma \approx \frac{-1+e}{\sqrt{2}}+ \frac{\sqrt{2}-1}{x} +
\mathcal{O}(\sqrt{x+\frac{1}{e}}) \sim \frac{\sqrt{2}-1}{x} \sim R^{-3} $. Of course in this limit $\beta \rightarrow 1$ one can explain the almost horizontal curve in the region of interval $R \in [10^{14}-10^{18}]cm$, although there is a small shift due to the fact that in our model $\beta = 1-\frac{1}{\Gamma^{2}}\approx 1 \frac{2x}{\sqrt{2}-1}$. For the non relativistic case where $x>>1$ or equivalently $z>>1$ or $R \in [10^{19}-10^{23}]cm$ in the region of an interest the $W_{0}(z)$ function has the following series expansion $W_{0}(z)\approx  lnz-lnlnz+\frac{lnlnz}{lnz}+o(\frac{lnlnz}{lnz})$ $\Gamma=-1+\frac{M_{0}}{m}productLog[\frac{m}{M_{0}}e^{2\frac{m}{M_{0}}-C}]$. Notice that in our model if $ x >> $, $\Gamma \approx 1- \frac{c+ln2}{x} \sim - \frac{c+ln2}{x} \sim R^{-3} $ (numerical result shows that $c<<-ln2$)
leading to $\beta \sim R^{-\frac{3}{2}}$ instead of $\beta \sim R^{-3}$ ref.\cite{13}-\cite{15}. This explains the difference slope ( in logarithmic scale between our model and that of ref. \cite{13}-\cite{15} In fact in our new model the slope is smaller than that of ref.\cite{13}-\cite{15}).

\begin{figure}[h!]
    \begin{center}
        \includegraphics[width=32pc]{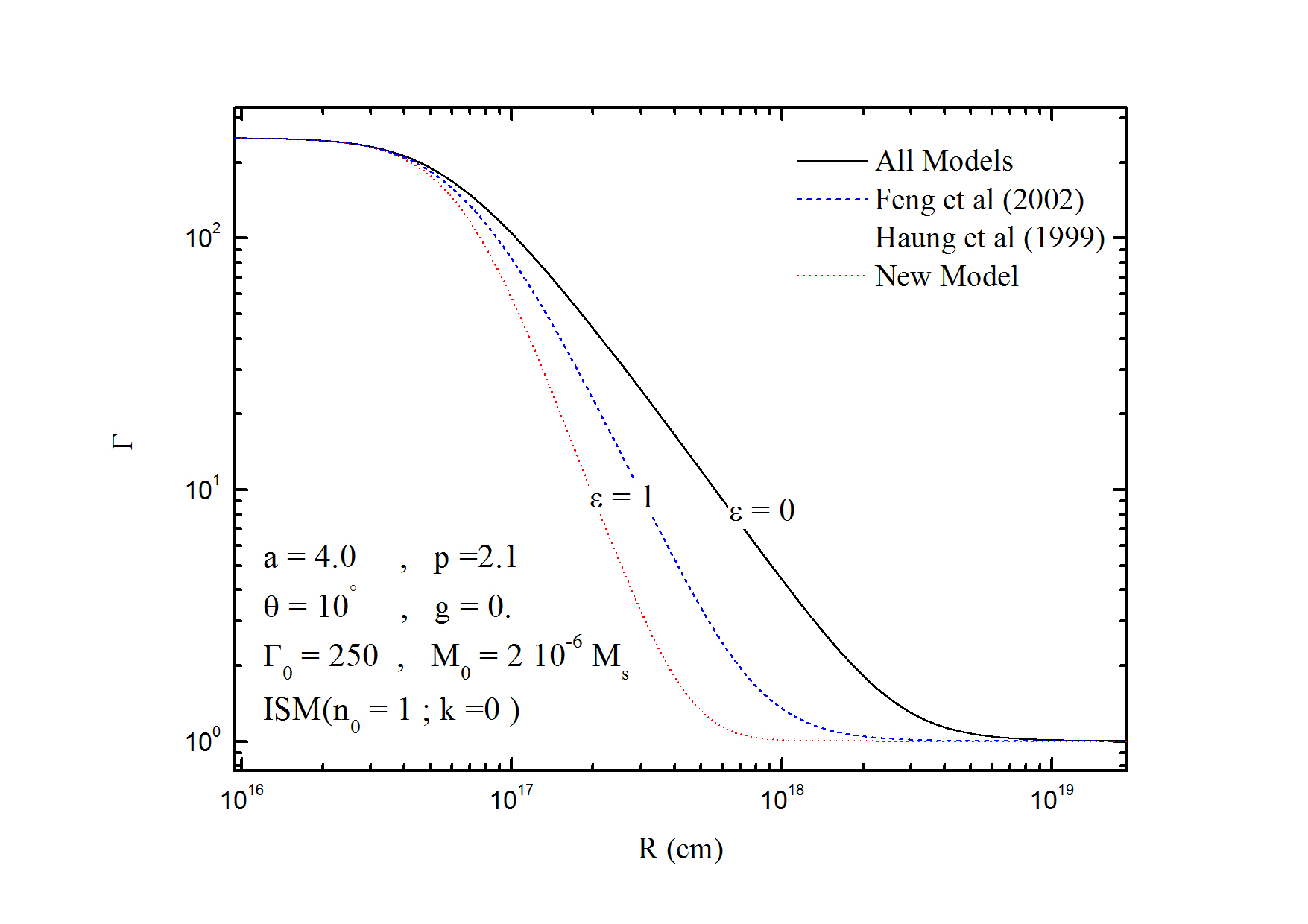}
    \end{center}
    \caption{\label{label} Evolution of the Lorentz factor $\Gamma$ as a function of the distance R (in logarithmic scale). For radiative ($\varepsilon=1$) and adiabatic ( $\varepsilon=0$) cases.}
    \label{fig:fa2}
\end{figure}

Similar behaviors are noticed in fig.\ref{fig:fa2} where the evolution of the Lorentz factor for the radiative case shows a faster drop (large slope) in our new model. Furthermore the curve of the three mentioned models coincide in the non-relativistic region.
In fact in fig.\ref{fig:fa2} (see the curve in UR limit), in the adiabatic case where $\epsilon=0$ and in all models we obtain the same differential  eq.\eqref{eq:18} which has as a solution eq.\eqref{eq:188} and therefore in the UR (resp NR) limit $\Gamma \sim R^{-\frac{3}{2}}$ (resp $\beta \sim R^{-\frac{3}{2}}$) since $m \sim R^{3}$. For the radiative case where $\epsilon=1$, in our model we end up with the differential equation $\frac{d\Gamma}{dm}=-\frac{\Gamma^{2}-1}{M_{0}+m(\Gamma+1)}$ which has as a solution $\Gamma=-1+\frac{M_{0}}{m}productLog[\frac{m}{M_{0}}e^{2\frac{m}{M_{0}}-C}]$. In the UR limit $\Gamma \approx (\sqrt{2}-1)\frac{M_{0}}{m}+\frac{e-1}{\sqrt{2}} \sim
\frac{M_{0}}{m} \sim R^{-3}$. However in the NR limit one gets $\Gamma \sim 1- (C+ln2)\frac{M_{0}}{m}$ ($\sim 1$) and $\beta \sim \sqrt{-2 \frac{C+ln2}{x}}\bigg/(1-\frac{C+ln2}{x}) \sim
x^{-\frac{1}{2}} \sim R^{-\frac{3}{2}}$.

It is worth to mention that, in the other models of ref. \cite{13}-\cite{11}, one has a different differential equation that is $\frac{d\Gamma}{dm}=-\frac{(\Gamma^{2}-1)}{M_{0}+m}$ which has as a solution $\frac{(\Gamma-1)(\Gamma_{0}+1)}{(\Gamma+1)(\Gamma_{0}-1)}=(\frac{m_{0}+M_{0}}{m+M_{0}})^{2}$ where $m_{0}$ is the initial mass. It is obvious that in the UR limit $\Gamma \sim \frac{M_{0}}{m}-1 \sim R^{-3}$ and in the NR limit $\beta \sim R^{-3}$ (difference in power of R in comparison to our model). Again, the difference in the slope of $\Gamma$ as a function of R (in a logarithmic scale ) is due essentially to the fact that we do not have the same solution of the two different differential equations mentioned above. Finally, in the NR limit, $\Gamma \rightarrow 1$ that is why all curves tends to the same limit.

\begin{figure}[h!]
    \begin{center}
        \includegraphics[width=30pc]{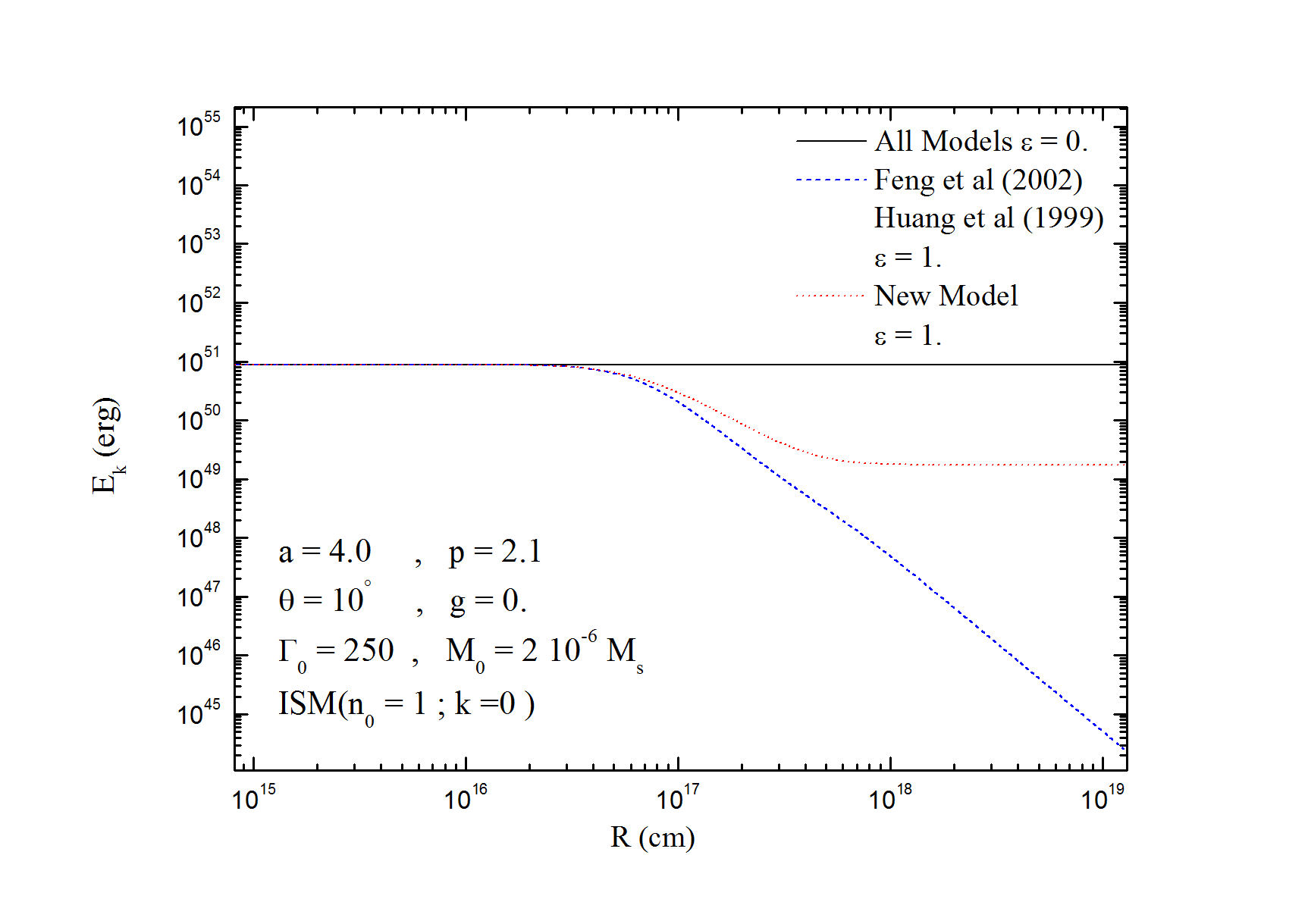}
    \end{center}
    \caption{\label{label} Evolution of the total  kinetic energy $E_{k}$ as a function of the distance R (in a logarithmic scale). For radiative ($\varepsilon=1$) and adiabatic ($\varepsilon=0$) cases.}
    \label{fig:fa3}
\end{figure}

Fig.\ref{fig:fa3} shows the total kinetic energy $E_{k}$ as a function of R. Notice that for the case $\varepsilon\simeq1.$ $E_{k}$ decreases slower in our new model than the proposed others models \cite{13} \cite{15} and it is almost flat in the range of $R\simeq10^{16}-10^{17}cm$ (UR) and $R\simeq10^{18}-10^{19}cm$ (NR). In the adiabatic case $E_{k}$ is constant for all the models $(E_{k}\sim10^{51}erg)$.

Fig.\ref{fig:fa3} displays the evolution of the total kinetic energy $E_{k}$ as a function of the distance R in a logarithmic scale for both the radiation and adiabatic cases where $\varepsilon = 1$ and $\varepsilon = 0$ respectively.
In fact we remaind that when $\varepsilon = 0$ for all models (including ours), the solution of eq.\eqref{eq:18} is that given by eq.\eqref{eq:188} leading to $E_{k}=E_{k_{0}}=(\Gamma-1)M_{0}c^{2}+(\Gamma^{2}-1)mc^{2}=C^{te}$ This explains why the kinetic energy is constant in dependently of the variation of R (both in the NR or UR cases). However, for $\varepsilon = 1$ (radiative case), the situation is different,especially in the NR case. The reason lies in the fact that eq. \eqref{eq:189} reduces to $\Gamma \sim 1-(C+ln2)\frac{M_{0}}{m}$  (as it was pointed out earlier) than the expansion of the kinetic energy $E_{k}=(\Gamma-1)M_{0}c^{2}$ (common to all models when $\varepsilon = 1$) will be $\sim -(C+ln2)\frac{M}{m}(M+M_{0})$ and since $m<<\gamma_{0}M_{0}$ than $E_{k}\sim -(C+ln2)=C^{te}$. For the other models, using eq.\eqref{eq:191} we deduce that $E_{k}\sim2\ \frac{(m_{0}+M_{0})^{2} (\Gamma_{0}-1) }{(\Gamma_{0}+1) }$ $
\frac{1}{m+M_{0}}\sim \frac{1}{m} \sim R^{-3}$. In the logarithmic scale, this behaves like a straight line with a slope $=-3$. For the UR case, our model gives using eq.\eqref{eq:191} and $ \Gamma \sim \frac{\sqrt{2}-1}{m}M_{0}\sim\frac{1}{m}\sim R^{-3}$, $E_{k}\sim\frac{\Gamma}{m+M_{0}}\sim\frac{\Gamma}{M_{0}}\sim
R^{-3}$. Similar expressions can be obtained for the other models. Notice in this case we have the same behavior of $E_{k}$ as a
function of R expect of course the proportionality factor a small
shift due the difference of the solution of $\Gamma$ between our model and the others.

\begin{figure}[H]
	\begin{center}
		\includegraphics[width=32pc]{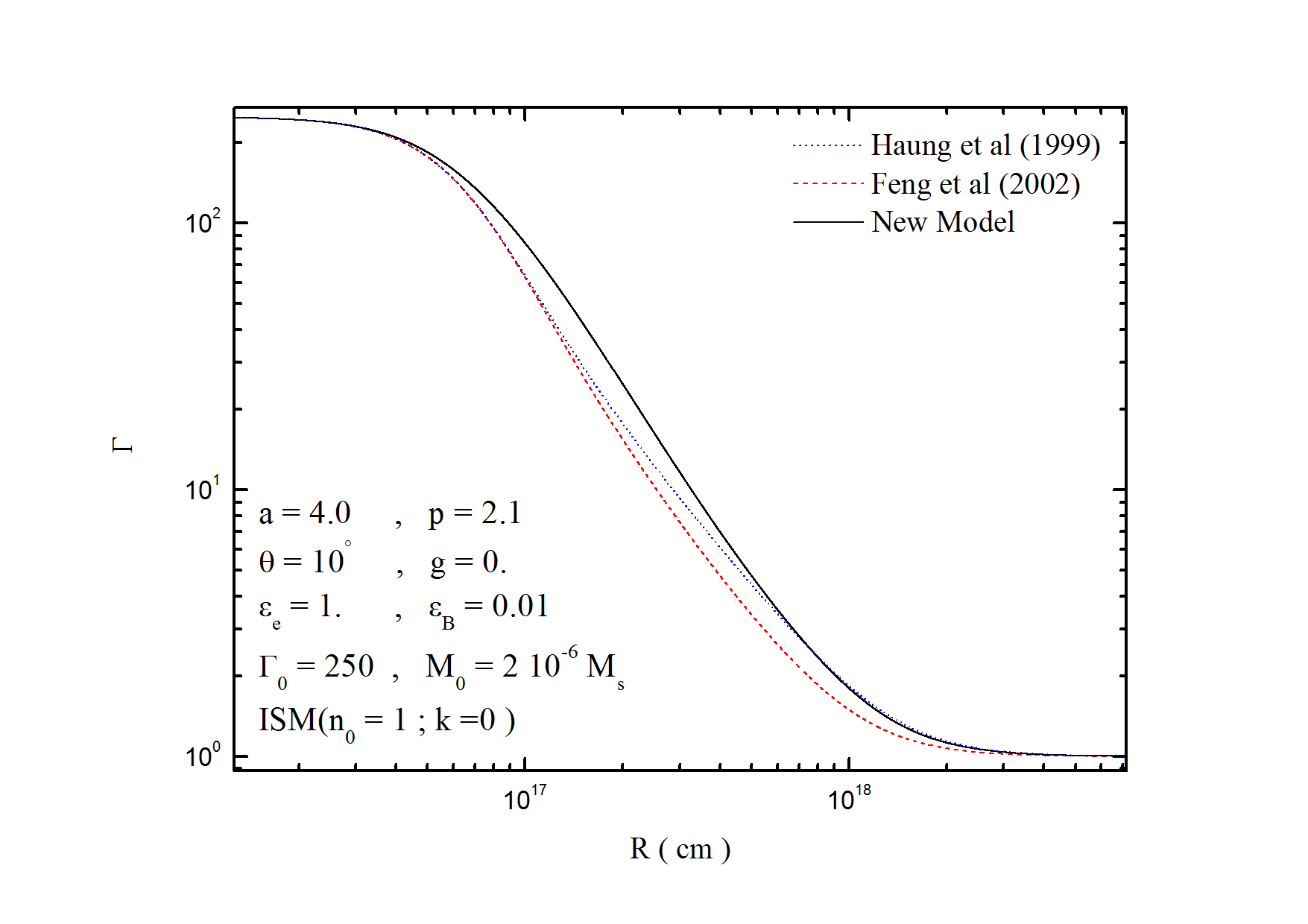}
	\end{center}
	\caption{\label{label} Evolution of the Lorentz factor $\Gamma$ as a function of the distance R (in logarithmic scale). For $\varepsilon_{e}=1.$ and $\varepsilon_{B}=0.01$.}
    \label{fig:fa4}
\end{figure}

\begin{figure}[h!]
    \begin{center}
        \includegraphics[width=30pc]{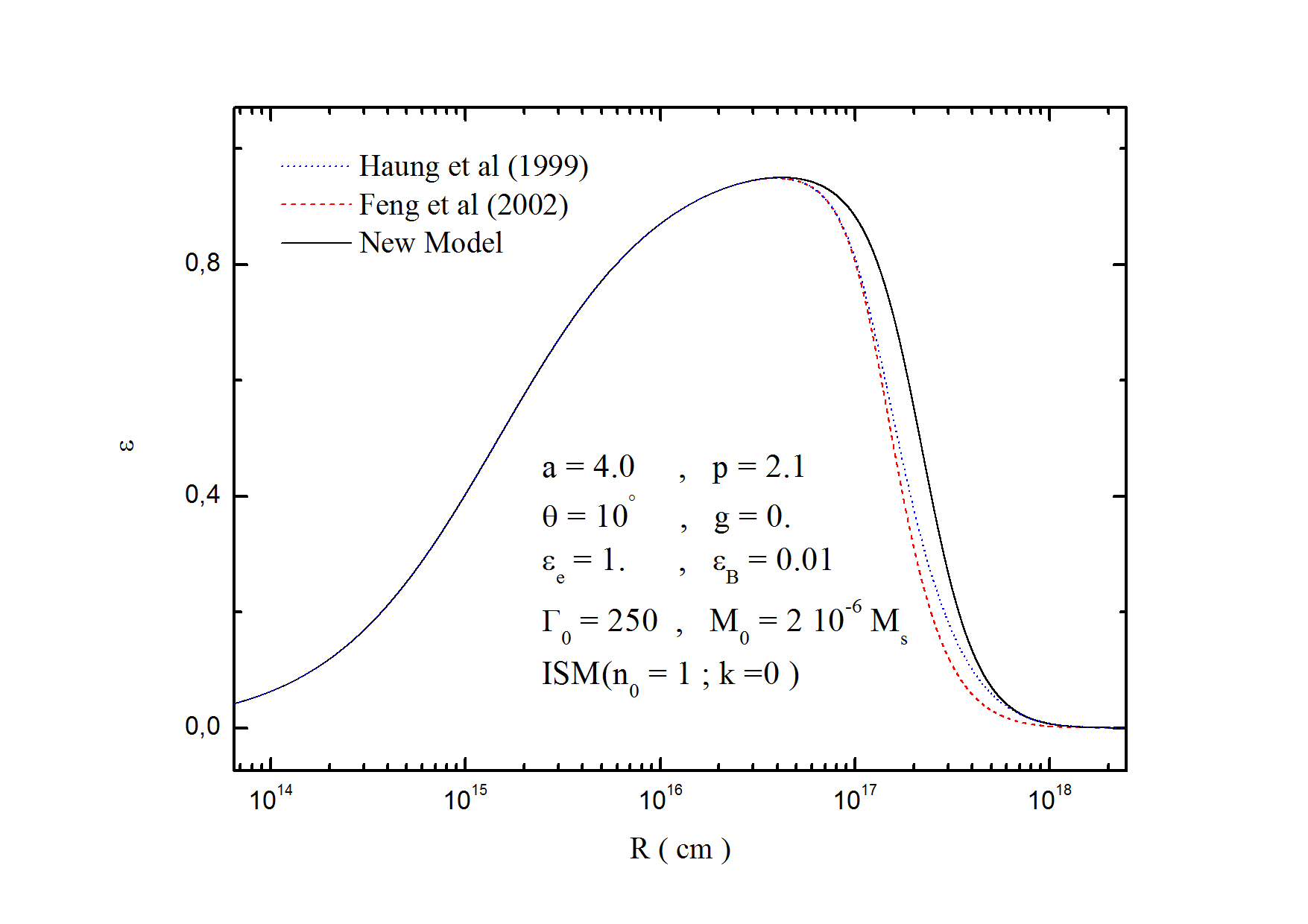}
    \end{center}
    \caption{\label{label} Evolution of the radiative efficiency of the fireball $\varepsilon$ as a function of the distance R (in logarithmic scale). For $\varepsilon_{e}=1.$ and $\varepsilon_{B}=0.01$.}
    \label{fig:fa5}
\end{figure}

\begin{figure}[H]
	\begin{center}
		\includegraphics[width=32pc]{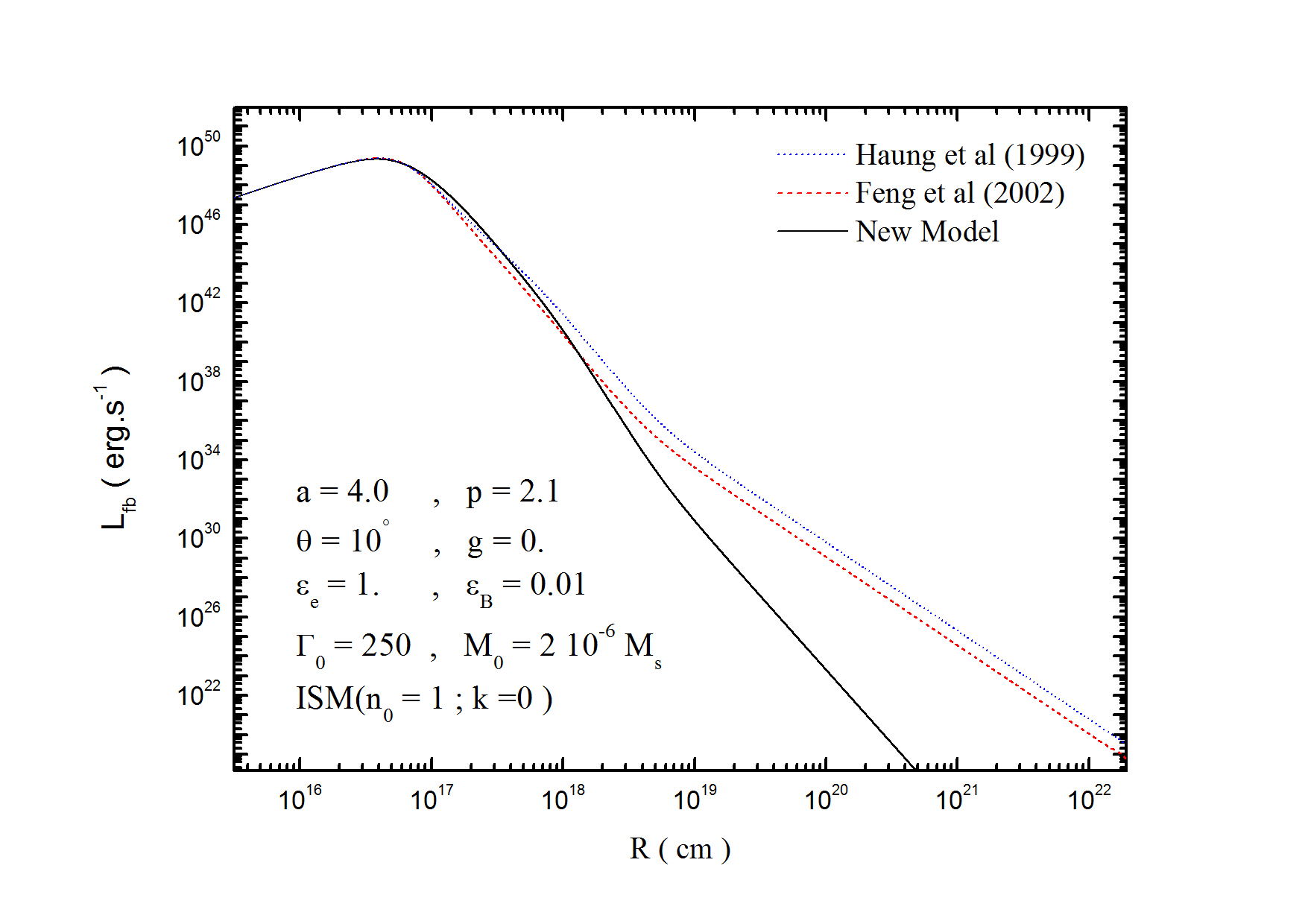}
	\end{center}
	\caption{\label{label} Evolution of the total luminosity of the fireball $L_{bf}$ as a function of the distance R (in logarithmic scale). For $\varepsilon_{e}=1.$ and $\varepsilon_{B}=0.01$.}
	\label{fig:fa6}
\end{figure}

\begin{figure}[H]
    \begin{center}
        \includegraphics[width=30pc]{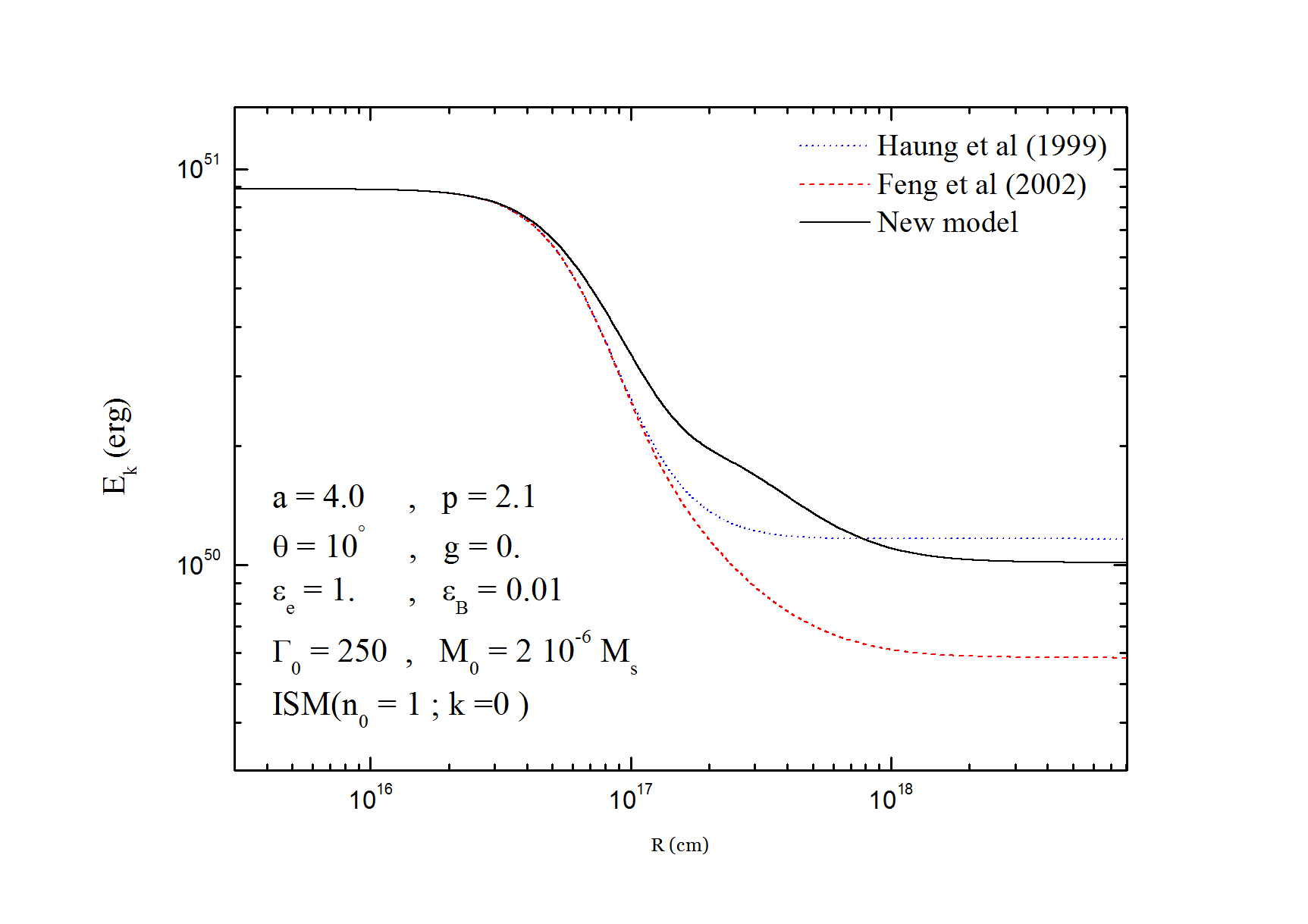}
    \end{center}
    \caption{\label{label}  Evolution of the total kinetic energy $E_{k}$ as a function of the distance R (in logarithmic scale). For $\varepsilon_{e}=1.$ and $\varepsilon_{B}=0.01$.}
    \label{fig:fa7}
\end{figure}

\begin{figure}[h!]
    \begin{center}
        \includegraphics[width=32pc]{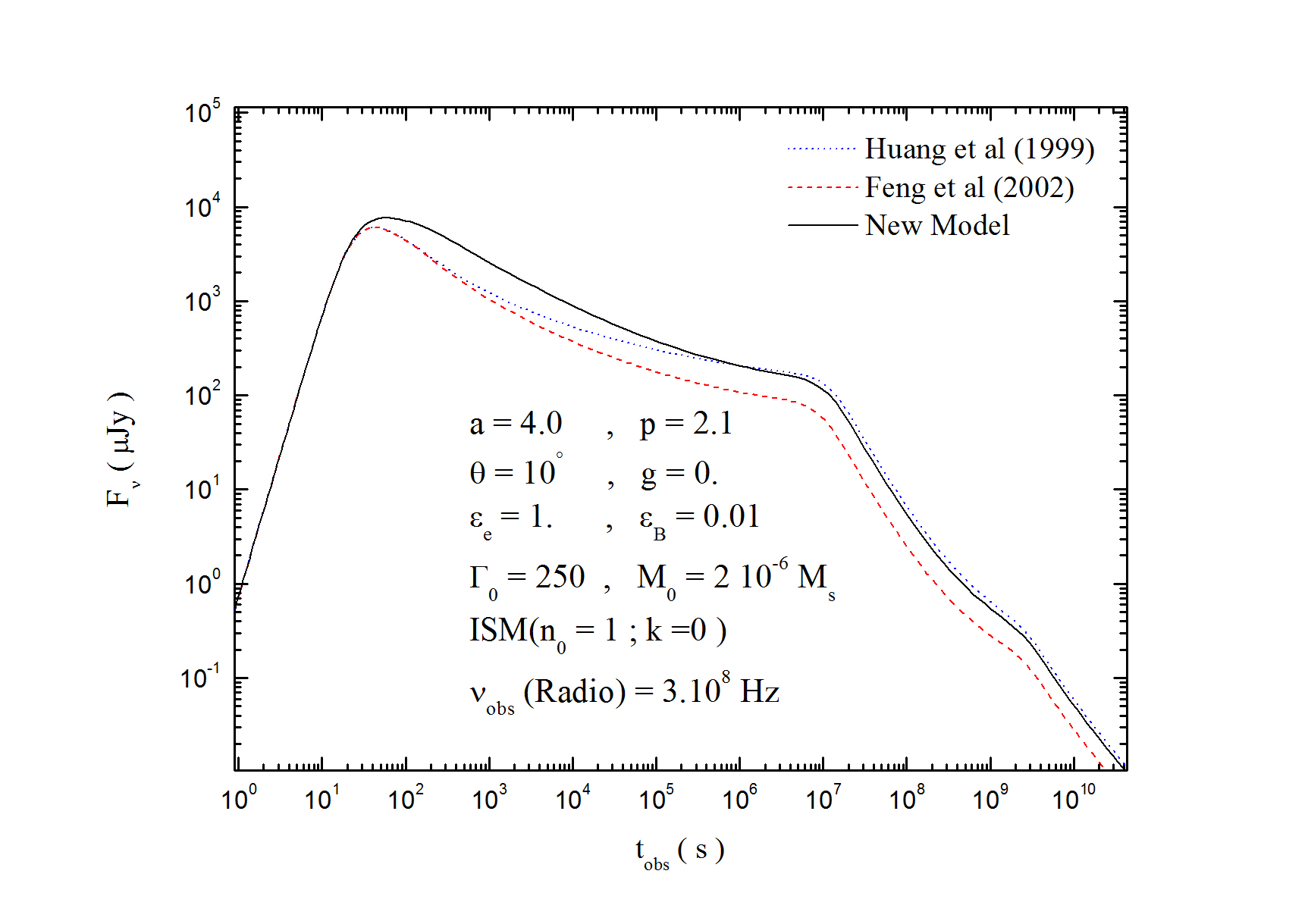}
    \end{center}
    \caption{\label{label} Light curve of GRB Afterglow (in logarithmic scale) within Huang et al (1999), Feng et al (2002) and the new models for radio frequency $ \nu_{obs}=3.10^{8}Hz$.}
    \label{fig:fa8}
\end{figure}

\begin{figure}[H]
    \begin{center}
        \includegraphics[width=32pc]{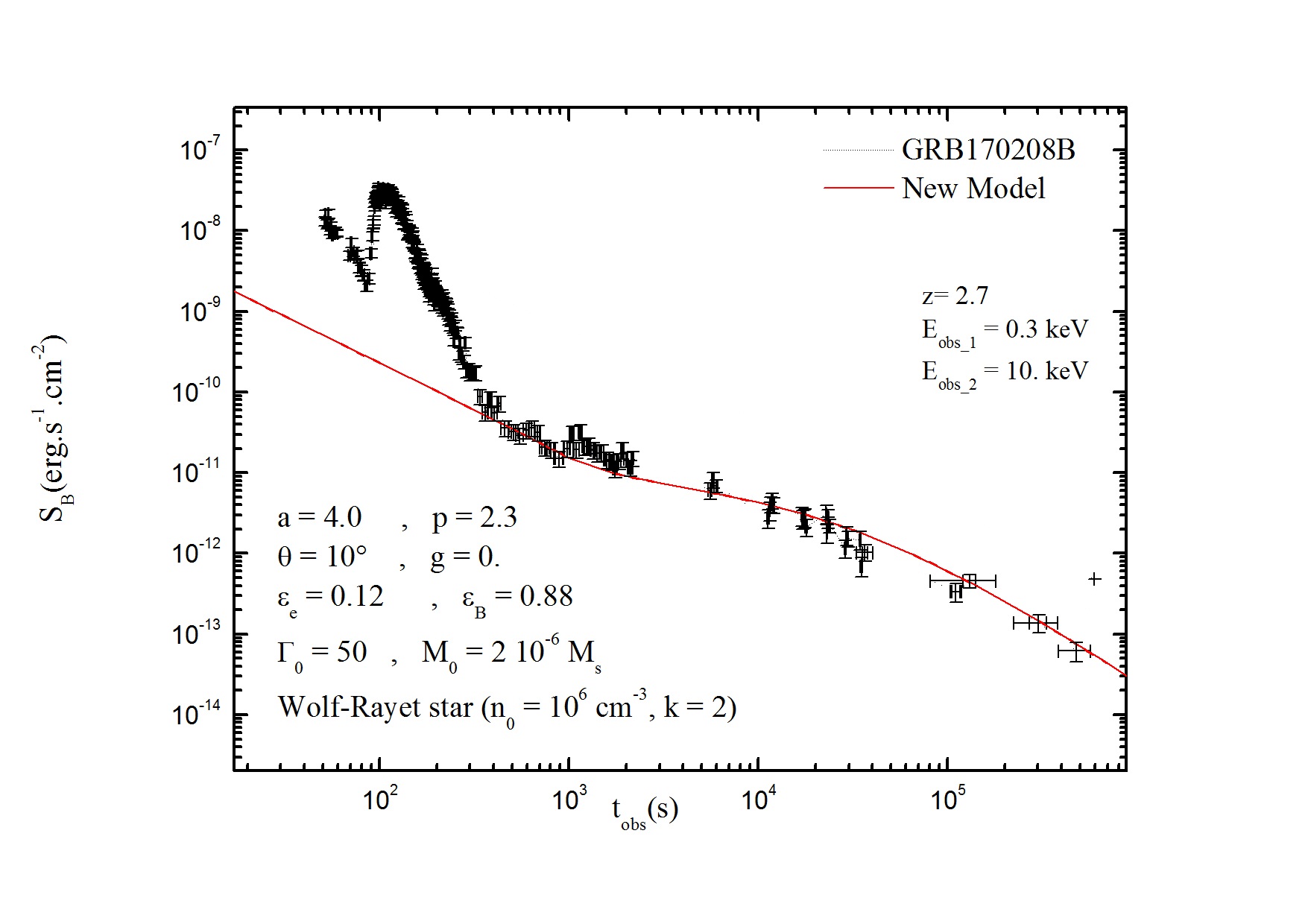}
    \end{center}
    \caption{\label{label} Integrated light curves of the GRB170208B Afterglow observed by the XRT / Swift telescope and their simulation (in logarithmic scale){\tiny }.}
    \label{fig:fa9}
\end{figure}

Fig.\ref{fig:fa5} shows the evolution of the radiative efficiency of the fireball $\varepsilon$ as a function of the distance R for our model (black solid line), Feng et al model (red dashed line)and
Huang et al (blue dashed line) for $\varepsilon_{e}=1.$ and $\varepsilon_{B}=0.01$. Notice that our curve is a little bit shifted in the interval $10^{17}-10^{18}cm$, and coincide in the region where $R>10^{18}cm$. Notice also in comparison with other models the maximum which appears at $R\sim5\times10^{16}cm$. For $R>>$ (NR case) the three curves coincide (similarly for the UR case).
For the intermediate values of R, the efficiency $\varepsilon$ in
our model is bigger than that of the Feng and Huang models for the
same radius R. For example for $R\sim 2 10^{17}$cm,
$\varepsilon_{our model}\approx 0.49$, $\varepsilon_{Huang}\approx
0.37$ and $\varepsilon_{Feng}\approx 0.27$. To justify this
behavior, we notice that using eqs. \eqref{eq:31} and \eqref{eq:32} defining $t^{'}_{syn}$ and $t^{'}_{ex}$ respectively, eq. \eqref{eq:33} takes the form $\epsilon= \varepsilon_{e} [1D \Gamma
R^{-1}(A\Gamma+B)^{-1}(\Gamma-1)^{-2}]^{-1}$ where A, B and D are constants.
Now in the NR region where R is large and $\Gamma \simeq 1$,
$\varepsilon \rightarrow \varepsilon_{e}=1$ (see our figure). However in the UR region where $\Gamma$ is larger $\varepsilon \rightarrow \varepsilon_{e}\frac{AR\Gamma^{2}\Gamma}{AR\Gamma^{2}+D}$ and this is a decreasing function of $R \Gamma^{2}$.Our numerical results show that $\frac{d\Gamma}{dt}=\frac{d\Gamma}{dR}\frac{dR}{dt}$ is a decreasing function of R and since $\frac{d\Gamma}{dt} \sim \Gamma^{2}$ (for a larger value of $\Gamma$ see eq.\eqref{eq:192}) then $\frac{d\Gamma}{dR}$ decreases faster that $\frac{1}{\Gamma^{2}}$ as a function of R and consequently $\varepsilon$ is a decreasing function of R as it is shown in fig.\ref{fig:fa5}. Now, the shift of $\varepsilon$ in our model compared to the others, lies in the fact that numerical results show that within the interval of $R \in [10^{16},10^{19}]$, the Lorentz factor $\Gamma$ of our model is greater than that of ref. \cite{13}-\cite{15} for a fixed value of R (see figure \ref{fig:fa4} ) or equivalently for a fixed value of $\Gamma$. The radius R of our model is greater than that of ref. ref. \cite{13}-\cite{15}. Therefore if eq.\eqref{eq:33} is rewriten as $\varepsilon = \varepsilon_{e} \frac{1}{D\Gamma / R(A\Gamma+B)(\Gamma-1)^{2}}$, than it is obvious that if $R_{our model}$ is greater than $R_{other models}$ (for a fixed value of $\Gamma$), one has $\varepsilon_{our model} > \varepsilon_{other models}$.

Fig.\ref{fig:fa6} displays the luminosity as a function of the radius R. Notice that there is a concordance between the three models in the range $10^{15}-10^{18}cm$ than the curve becomes steeper in our model in the range $>10^{19}cm$.

Fig.\ref{fig:fa6} shows the luminosity $L_{bf}$ as a function of R for a variable efficiency $\varepsilon$. Notice that L is a decreasing function of R and for relatively smaller values of R (UR case) it is almost of the same order (for our model and those of ref. \cite{13} \cite{15}) with a small shift difference (ours is larger). For example for $R=10^{16}cm$, $L_{our model} \approx 1.44 10^{49} erg.s^{-1}$ and $L_{Feng} \approx L_{Huang} \approx 9 10^{48} erg.s^{-1}$. Now, if R increases, $L_{our model}$ decreases faster than $L_{Feng}$ and $L_{Huang}$ for example for $R=10^{19}cm$, $L_{our model} \approx 10^{30} erg.s^{-1}$ and $L_{Feng} \approx 3.13 10^{33} erg.s^{-1}$ and $L_{Huang} \approx 1.92 10^{34} erg.s^{-1}$. Therefore, for larger values of R (NR case), the discrepancy increases between the results of our model and those of ref. \cite{13} \cite{15}.This is due mainly to the fact that L contains two compelling terms $L_{1}=\Gamma \varepsilon (\Gamma-1)\frac{dm}{dt}$ which is positive and $L_{2}=\Gamma
\varepsilon m \frac{d\Gamma}{dt}$ (present only in our model), which is negative. As R increases, $L_{2}$ increases and compensates $L_{1}$, such that $L=L_{2}+ L_{1}$ become smaller in comparison with the other models where the term $L_{2}$ is absent. Table~\ref{tab :ta1}, summarizes some numerical values illustrating this fact.

\begin{table}[H]
	\begin{center}
		\begin{tabular}{|c|c|c|c|c|c|}
			\hline                     &                        \multicolumn{3}{c|}{$L_{1}(erg.s^{-1})$}                   &    $L_{2}(erg.s^{-1})$      &  $L_{1}+L_{2}(erg.s^{-1})$  \\
			\hline       R(cm)         &          Huang          &            Feng           &                      \multicolumn{3}{c|}{Our model}                                     \\
			\hline $1.09\times10^{16}$ &  $3.48785\times10^{48}$ &   $3.48786\times10^{48}$  &   $3.48798\times10^{49}$    &     $-1.5353\times10^{46}$  &  $3.47263\times10^{48}$     \\
			\hline $1.13\times10^{17}$ &  $5.5\times10^{47}$     &   $4.29\times10^{47}$     &        $2.09\times10^{48}$  &     $-1.16\times10^{48}$    &  $9.36\times10^{47}$        \\
			\hline $1.03\times10^{18}$ &  $2.5\times10^{41}$     &   $2.0\times10^{40}$      &     $1.71\times10^{41}$     &     $-1.40\times10^{41}$    &  $3.08\times10^{40}$        \\
			\hline $1.06\times10^{19}$ &  $1.92\times10^{34}$    &   $3.13\times10^{33}$     &     $1.3323\times10^{34}$   &     $-1.3318\times10^{34}$  &  $5.01\times10^{30}$        \\
			\hline $1.10\times10^{20}$ &  $4.29\times10^{29}$    &   $7.63\times10^{28}$     &    $3.03813\times10^{29}$   &    $-3.03812\times10^{29}$  &  $1.04\times10^{23}$        \\
			\hline
		\end{tabular}
		\caption{Some numerical results illustrating the luminosity of the three models}
		\label{tab :ta1}
	\end{center}
\end{table}

Fig.\ref{fig:fa7} displays the total kinetic energy $E_{k}$ as a function of the distance R. For $\varepsilon_{e}=1.$ and
$\varepsilon_{B}=0.01$. Notice the discrepancy between the three models for the range $>5\times10^{16}cm$. The curves in the three models became almost flat in the region $R>10^{18}cm$. Notice also the existence of tow slopes in our new model in the region $R>5\times10^{16}-10^{18}cm$.

Fig.\ref{fig:fa7} displays $E_{k}$ as a function of R. Notice that for values of $R=10^{16}cm$ (UR region), $E_{k}$ takes almost the same value for the three models ($E_{k} \approx 8.86 \times 10^{50} erg$). The reason lies to the fact that in the UR region where $\Gamma$ is larger and $M_{0} \gg m$, $\varepsilon \approx \varepsilon_{e}$ (see eq.\eqref{eq:33}) and therefore in all models we have $E_{k} \sim (\Gamma-1) M_{0}$ and almost the same differential equation $\frac{d\Gamma}{dm}=-\frac{\Gamma^{2}-1}{M_{0}}$. In the NR region where $m \gg M_{0}$ and $\Gamma \approx 1$, $E_{k}$ becomes constant. For example for  $R \approx 10^{19} cm$, $E_{k(Huang)} \approx 1.16 \times 10^{50} erg$, $E_{k(Feng)} \approx 6.09 \times 10^{49} erg$ and $E_{k(Our model)} \approx 1.07 \times 10^{50} erg$ and $R \approx 10^{20} cm$, $E_{k(Huang)} \approx 1.15 \times 10^{50} erg$, $E_{k(Feng)} \approx 5.79 \times 10^{49} erg$, $E_{k(Our model)} \approx 1 \times 10^{50} erg$, (see table~\ref{tab :ta2}). The reason lies to the fact that, in the NR region where $\Gamma \approx 1$, $\varepsilon \approx 0$ and $U \approx m(\Gamma-1)+c^{te}$, one has $\frac{d\Gamma}{dm} \sim - \frac{\Gamma^{2}-1}{M_{0}}$ for all models  which leads to $m \sim \frac{1}{\Gamma-1}$ and therefore $E_{k} \approx 2(\Gamma-1)m+C^{te}$ becomes constant. The difference in the constant between the three models is due to the complicated expression of $\varepsilon \approx \varepsilon_{e} \times 1/[R(A\Gamma+B)(1-\Gamma)^{2}]$ (if $\Gamma \rightarrow 1$, where $\Gamma$ is a function of m or equivalently R because slightly different our model to an other. Any way, numerical results shows that in the NR case, $E_{Huang}$ becomes slightly bigger than ours and almost twice that of Feng and al. In the intermediate region where $R \approx 3 \times 10^{17} cm$; $E_{Our model} \approx 1.68 \times 10^{50} cm$, $E_{Feng} \approx 8.81 \times 10^{49} cm$ and $E_{Huang} \approx 1.22 \times 10^{50} cm$. In fact, the difference between our model and Feng et al \cite{15} (resp. Huang et al \cite{13}) is due to the difference in the expression of the used differential equation $\frac{d\Gamma}{dm}$ (see eqs. \eqref{eq:2} and \eqref{eq:3}) (resp. to the difference in the experience of $E_{k}$ (see eq.\eqref{eq:5})and U (see eq.\eqref{eq:2343} and $U=U_{ex}=(\Gamma-1)mc^{2}$). Table ~\ref{tab :ta2}, summarizes the behavior of $E_{k}$ in various regions.

\begin{table}[H]
	\begin{center}
		\begin{tabular}{|c||c|c|c|}
			\hline                          &                    \multicolumn{3}{c|}{$E_{k}$ (erg)}                           \\
			\hline        R(cm)             &           Huang          &          Feng           &         Our model       \\
			\hline \hline   $10^{16}$         &   $8.86 \times 10^{50}$  &  $8.86 \times 10^{50}$  &  $8.86 \times 10^{50}$  \\
			\hline        $10^{17}$         &   $2.63 \times 10^{50}$  &  $2.59 \times 10^{50}$  &  $3.40 \times 10^{50}$  \\
			\hline      $3 \times 10^{17}$  &   $1.22 \times 10^{50}$  &  $8.81 \times 10^{49}$  &  $1.68 \times 10^{50}$  \\
			\hline        $10^{18}$         &   $1.16 \times 10^{50}$  &  $5.09 \times 10^{49}$  &  $1.07 \times 10^{50}$  \\
			\hline        $10^{19}$         &   $1.16 \times 10^{50}$  &  $5.83 \times 10^{49}$  &  $1.01 \times 10^{50}$  \\
			\hline        $10^{20}$         &   $1.15 \times 10^{50}$  &  $5.79 \times 10^{49}$  &  $1.00 \times 10^{50}$  \\
			\hline
		\end{tabular}
		\caption{Some numerical results illustrating the kinetic energy of the three models}
		\label{tab :ta2}
	\end{center}
\end{table}

\begin{table}[H]
	\begin{center}
		\begin{tabular}{|c||c|c|c|}
			\hline                      &        \multicolumn{3}{c|}{$F_{\nu}$ ($\mu Jy$)}            \\
			\hline    $t_{obs}$(s)      &       Huang       &     Feng      &     Our model   \\
			\hline \hline     10        &      789.42       &    789.44     &       789.98    \\
			\hline            43        &       6045        &     6049      &       7333      \\
			\hline $1.2 \times 10^{3}$  &       1112        &     890       &       3159      \\
			\hline        $10^{6}$      &       205         &     98        &       207       \\
			\hline        $10^{8}$      &       6           &     1         &       4         \\
			\hline        $10^{9}$      &       0.6         &     0.2       &       0.4       \\
			\hline        $10^{10}$     &       0.05        &     0.01      &       0.04      \\
			\hline
			\end{tabular}
			\caption{Some numerical results illustrating the instantenos intensity of light curve of the three models}
		\label{tab :ta3}
	\end{center}
\end{table}

Fig.\ref{fig:fa8} shows the light curve of GRB Afterglow within all model for a radio frequency $\nu_{obs}=3.10^{8}Hz$. Notice the same behavior of the three models with a small shift in comparison with Feng when $R\leq10^{2}cm$. Notice also that:

\begin{enumerate}[1)]
	\item $F_{\nu}$ is a decreasing function of t as it is expected.
	\item There are five regions where the slope $\frac{dF_{\nu}}{dt}$ change its direction. This is due to the complicated behavior of $\frac{dP}{dt}$ as a function of t.
	\item There is a maximum (for all models) around $t \sim 43s$ ($F_{\nu (Huang)}	\approx 6045 erg.s^{-1}.cm^{-2}, F_{\nu (Feng)} \approx 6049 erg.s^{-1}.cm^{-2}, F_{\nu (Our model)} \approx 7333 erg.s^{-1}.cm^{-2}$). This can be explained by the fact that the	synchrotron function K(x) has a maximum around $x \approx 0.3$ and	therefore $F_{\nu}$ should has a maximum corresponding approximately to $t \approx 43s$
	\item For relatively small values of t (region 1, $t \sim 10s$), one has almost the same values of $F_{\nu}$ for qll models. That is: $F_{\nu (Huang)} \approx 789.42 erg.s^{-1}.cm^{-2}, F_{\nu (Feng)} \approx 789.44 erg.s^{-1}.cm^{-2}, F_{\nu (Our model)} \approx 789.98 erg.s^{-1}.cm^{-2}$. 
	\item In the region 2 (where $\frac{dF_{\nu}}{dt}$ change the 
	direction), our $F_{\nu}$ is above and different from that of Huang et al and Feng et al. For example for $t \sim 1.2 \times 10^{3}s$	$F_{\nu (Huang)} \approx 1112 erg.s^{-1}.cm^{-2}, F_{\nu (Feng)} \approx 890 erg.s^{-1}.cm^{-2}, F_{\nu (Our model)} \approx 3159 erg.s^{-1}.cm^{-2}$
	\item In the region 3 where $t \sim 10^{6}s$ and $F_{\nu (Huang)} \approx 250	erg.s^{-1}.cm^{-2}, F_{\nu (Feng)} \approx 98 erg.s^{-1}.cm^{-2} and F_{\nu (Our model)} \approx 207 erg.s^{-1}.cm^{-2}$, $\frac{dF_{\nu}}{dt}$ changes again its direction (decreases). 
	\item In the region 4 where $F_{\nu}$ drops faster (a more pronounced slope is noticed) $t \sim 10^{8}-10^{9}s$ and $F_{\nu (Huang)} \in [6,0.6] erg.s^{-1}.cm^{-2}, F_{\nu (Feng)} \in [1,0.2]	erg.s^{-1}.cm^{-2} and F_{\nu (Our model)} \in [0.4,0.04]	erg.s^{-1}.cm^{-2}$.
	\item In the region 5 where $t \succeq 10^{10}s (R  \approx 10^{19}cm)$ and $F_{\nu (Huang)} \preceq 0.05 erg.s^{-1}.cm^{-2}, F_{\nu (Feng)} \preceq 0.01 erg.s^{-1}.cm^{-2}$ and $F_{\nu (Our model)} \preceq 0.04 erg.s^{-1}.cm^{-2}$, $F_{\nu}$ changes again its slope (increases).
\end{enumerate}

Notice that in all region 1,2 and 3 in our model $F_{\nu}$ is above and almost the same as that of Huang et al but slightly different (almost a constant) from that of Feng et al.

Fig.\ref{fig:fa9} displays the light curves of the GRB170208B Afterglow observed by the XRT / Swift telescope and their simulation. Notice the fairly good agreement of our model in the X-Ray Afterglows region $t_{obs} \succeq 3\times10^{2}s$ and this confirm the viability of our model.

\section{Conclusion}
Throughout this paper, a new generic model of the GRB Afterglows which verifies the Sedove solution and energy conservation in the adiabatic regime is constructed. We have studied and discussed the evolution of the jet velocity, Lorentz factor, kinetic energy, efficiency radiation coefficient, luminosity and instantenous intensity of light curve as a function of the radius of the blast wave R and$/$or time t.
We have considered both the radiative and adiabatic cases as well as constant and variable efficiency radiation coefficient $\varepsilon$. The results f our model are compared with those of Huang et al and Feng et al models. It turns out that the behaviors and$/$or numerical results are almost the same in the ultra relativistic region and different globally in the other region (relativistic and non relativistic). Finally, fairly good concordance with the observational data of XRT$/$Swift telescope concerning the integrated light curves of the GRB170208B was obtained in the interval of our interest  $t_{obs} \succeq 3\times10^{2}s$.

\section{Acknowledgments}
We are very grateful to the Algerian Minister of Higher Educations and Scientific Research and DGRET for the financial support.

\end{document}